\documentclass[aip,jcp,preprint]{revtex4-1}
\usepackage{graphicx}
\usepackage{xcolor}
\usepackage{amsmath}
\usepackage{amssymb}
\usepackage{fixmath}
\newcommand{\cF}{\mathcal{F}}
\newcommand{\hc}{\mathrm{hc}}
\newcommand{\el}{\mathrm{el}}
\begin{document}

\title{Ionic Size Effects on the Poisson-Boltzmann Theory}

\author{Thiago Colla}
\email{colla@iceb.ufop.br}
\affiliation{Instituto de F\'isica, Universidade Federal de Ouro Preto, CEP 35400-000, Ouro Preto, MG, Brazil}

\author{Lucas Nunes Lopes}
\affiliation{Instituto de F\'isica, Universidade Federal do Rio Grande do Sul, Caixa Postal 15051, CEP 91501-970, Porto Alegre, RS, Brazil.}

\author{Alexandre P. dos Santos}
\email{alexandre.pereira@ufrgs.br}
\affiliation{Instituto de F\'isica, Universidade Federal do Rio Grande do Sul, Caixa Postal 15051, CEP 91501-970, Porto Alegre, RS, Brazil.}
\affiliation{Fachbereich Physik, Freie Universit\"at Berlin - 14195 Berlin, Germany.}

\begin{abstract}
In this paper we develop a simple theory to study the effects of ionic size on ionic distributions around a charged spherical particle. We include a correction to the regular Poisson-Boltzmann equation in order to take into account the size of ions in a mean-field regime. The results are compared with Monte Carlo simulations and a Density Functional Theory based on the Fundamental Measure approach and a second-order bulk expansion which accounts for electrostatic correlations. The agreement is very good even for multivalent ions. Our results show that the theory can be applied with very good accuracy in the description of ions with high effective ionic radii and low concentration, interacting with a colloid or nanoparticle in an electrolyte solution.
\end{abstract}

\maketitle

\section{Introduction}
 
Many complex systems of paramount relevance in the fields of biology and physical chemistry can be to some extent mapped into a simplified picture in which nanoparticles of different sizes and shapes coexist with smaller components in a solvent environment. In many practical applications, these particles are constrained to move in narrow regions whose typical sizes have range of magnitudes not too far from the particles suspended in it~\cite{Mes09}. A classical example is the compact environment inside the inter-cell space, where many different species coexist. In these crowded environments, exclusion volume effects between the different constituents play a key role in determining the majority of system properties~\cite{Min92,Min95,Zim93,Lu11,Phi13,Lim14,Lim16_1,Lim16_2}. In some cases the nanoparticles can be permeable to some of the smaller ones, in which case exclusion volume effects will strongly influence the osmotic flow at the nanoparticle interface, eventually leading to particle size fluctuations \cite{YLev02,Col14,Den16}. Apart from such volume changes, soft nanoparticles can also display strong shape deformations due to their exclusion volume interactions with the different system components -- giving rise to complex and rich equilibrium particle topologies that are dictated by both size and concentrations of the smaller components \cite{Mat16,Rie15,Lim14}. Hard nanoparticles are on the other hand not allowed to fluctuate in neither shape nor size. These constraints force particles to be arranged in ordered structures whenever the overall packing fraction is sufficiently high. This is not only the case of the aforementioned molecular crowding, but might also occur for instance in nano-confined solvents or ionic liquids. Such ordered structures are clearly a result of strong positional correlations among the hard particles. The physical mechanisms behind these packing effects go way beyond the classical interpretation based on the entropy loss induced by exclusion volume effects -- as reflected by the reduced space available for particles to diffuse in. Indeed, an accurate description of such correlation-induced effects in hard systems requires the use of sophisticated approaches, generally non-local in nature \cite{Reis59,Leb65,Ros90,Stil06}. Even standard molecular simulation techniques have to be adapted in order to circumvent problems associated with the small particle mobility resulting from frequent particle collisions \cite{Kur11}. In spite of this underlying complexity in particle structure, it is important to emphasize that still many important properties of polydisperse hard systems can be pretty well understood on the basis of the simplified picture of entropy reduction resulting from exclusion volume effects \cite{Lik01}, whose theoretical description dates back to the pioneering work of Asakura and Oosawa \cite{Asa58}. It is the case for instance of the well-known attraction between nanoparticles driven by depletion effects upon addition of smaller hard components \cite{Tui03,Dzu90,Cam10,Cam12,Han00}, which is very important in a number of applications involving nanoparticle stabilization \cite{Lik01,Bel00,Roij03}. 

The system complexity increases even further when, apart from finite size effects, electrostatic interactions become also relevant. This is usually the case when the nanoparticles are suspended in an aqueous solvent, which favors the partial dissociation of ionic groups at their surfaces~\cite{Lev02}. Addition of salt is also a possibility, in general aimed to avoid irreversible particle aggregation driven by the aforementioned depletion attractions and Van der Waals Forces. For hard nanoparticles, the strong electrostatic correlations will result in a large number of small ions surrounding their surfaces to form a charge structure which is widely known as the Electric Double Layer (EDL)~\cite{Lev02,Han00}. In some situations, such ionic adsorption at the nanoparticle surface will be crucial to determine the ionic induced, effective interactions among them~\cite{Lev02,Bel00,Klein02,Dob06}. Size effects, on the other hand, will limit the number of such smaller components that can be assembled at the vicinity of the charged surfaces. Depending on the strength of ion-ion electrostatic interactions, positional correlations driven by these interactions can also become very relevant~\cite{Lev02}. Many interesting (and sometimes quite counter-intuitive) phenomena can appear as a consequence of electrostatic correlations~\cite{Loz82,Val91,Ter01,Que03,Lev04,Pia05,Diehl06,Mes09}, the theoretical description of which is a difficult task. How relevant these correlations will be depends on a fine tuning between ionic charge, size and the dielectric constant of the solvent they are embedded into~\cite{Lev02}. The relative strength of electrostatic correlations can be quantified via the so-called {\it coupling parameter}~\cite{Lev96,Naij05,Pai11,San16}, which measures the ratio between the electrostatic energy at ionic contact and the ionic thermal energy. Even if the ions bear relatively large charges (e .g.  multivalent ions), their size can be large enough such as to prevent a close center-to-center approach, thereby limiting the strength of the electrostatic interactions. Finite size correlations should then play the major role in determining the structure of the EDL. 

When the ionic coupling is not too high, the electrostatic correlations are in general weak enough to be safely neglected. In such cases, the mean-field Poisson-Boltzmann~(PB) theory is known to accurately describe the EDL structure, provided the ionic components are not too large in size. Since the mean-field theory is originally designed to deal with point-like ions, it is not able to properly distinguish electrostatic forces among charged particles having different sizes. Therefore, in situations of large ions or at high electrostatic couplings, the theory is no longer able to capture the EDL features resulting from strong ionic correlations, in such a way that more elaborated theories -- such as the Density Functional Theory~(DFT)~\cite{Evans79,Blum81,Ros93,Li04,Yu04,Yan15}, Integral Equations theory~\cite{Att96,Hen78,Hen79,Ver82,Ball86,Kje86,Kje88,Kje88_2,Kje92} or theoretical field approaches~\cite{Mor00,Mor19,Naij05} -- have to be employed. Although these theories go far beyond the range of validity of the traditional PB approach, they sometimes lack in physical transparency and in most cases their implementation is far from straightforward. In order to keep the simplici\-ty and numerical efficiency inherent to the PB approach --  and yet be able to accurately describe systems with non-negligible ionic correlations -- many attempts have been made over the years towards the direction of extending the mean-field approach to incorporate both size~\cite{Lev60,Bor97,Bor00,Boh01,Ant05,Bhu09,Lou09,LG11} and electrostatic~\cite{Bar00,San09,San10,Col10,Bakh11} correlation effects. This has resulted in what is generally known as modified Poisson-Boltzmann~(mPB) approaches. Many of these modifications provide indeed quite an improvement over the mean-field predictions, although they range of validity is usually restricted to some specific situations that have to be further tested against more refined approaches. 

Quite recently, one such mPB has been designed by dos Santos \textit{et al.}~\cite{DoBa16} in order to account for finite size effects on the adsorption of charged hard nanoparticles at a charged interface. The model is based on a local approximation that renormalizes the screening length taking proper account for the finite colloidal size. The predictions for the distribution of charged nanoparticles near the interfaces were compared with Monte Carlo simulations in the dilute regime, showing a very good agreement~\cite{DoBa16}. Since the numerical simulation of concentrated charged nanoparticles is a very challenging task, it is not clear whether the model will be accurate for concentrated hard systems as well, as the finite size effects become very strong. In the present work, we aim to adopt a similar model to describe the distribution of polydisperse hard ions around a charged spherical colloid. Instead of a planar geometry, we will adopt a spherical cell~(SC) model, suited for concentrated colloidal dispersions~\cite{Des01}. The SC model allows one to compare simulation and theory from moderate to high concentrations, providing a perfect framework to test the range of validity of the underlying theoretical assumptions. Besides, we will proceed to investigate the relative influence of finite size effects on both hard-core and electrostatic correlations in suspensions containing multivalent ions of different sizes. To this end, we will apply a DFT that combines hard sphere and electrostatic correlation effects through the Fundamental Measure Theory~(FMT) and a second-order bulk expansion, respectively. The DFT allows one to easily control different correlations in order to investigate the regimes in which their relative strengths become more relevant.
  
The paper is structured as follows. In the next section, the system under consideration is described in some detail. The mPB that incorporates finite ionic sizes into the PB framework is presented hereafter. Next, we summarize the DFT approach to be applied, as well as the MC techniques used to obtain the ionic density profiles. Finally, predictions from the different approaches for the ionic distributions in the context of the SC cell model are shown and discussed in detail. Conclusions and perspective for future applications are then outlined in the last section. 

% The present idea was recently applied to the study of adsorption of colloids to a charged wall~\cite{DoBa16}.

\section{Model system}\label{secII}

\begin{figure}[h!]
\begin{center}
\includegraphics[width=6cm]{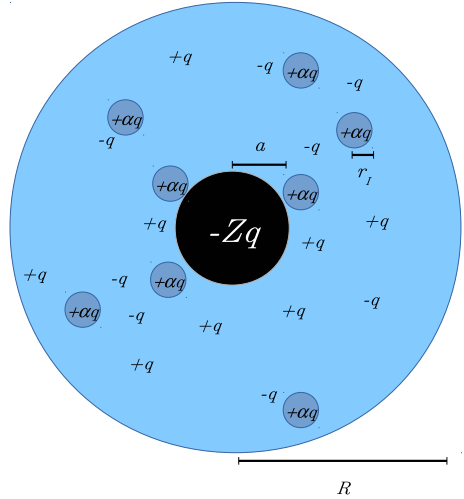}
\end{center}
\caption{Schematic representation of the system under investigation. A colloid of charge $-Zq$ and radius $a$ is centered in a confining SC of radius $R$. Ions of charges $\pm q$ and effective radius $r_i$ (length not represented in this view), and ions of charges $+\alpha q$ and effective radius $r_I$ are also present in solution.}
\label{fig1}
\end{figure}

We adopt a SC model, a sphere with radius $R$, see Fig.~\ref{fig1}. The colloid is represented by a sphere of radius $a$ whose center is located at the origin, bearing a uniformly distributed negative surface charge $-Zq$, where $q$ is the proton charge. The radius of colloid is set to $a=50~$\AA, while its charge is $Z=60$. Besides $Z$ positive counterions, modeled as hard spheres with radius $r_i$ and charges $+q$,  there are in addition ions from dissociation of two different salts. The counterions from dissociation of asymmetric $\alpha:1$ salt are modeled as hard spheres with radius $r_I$ and charge $+\alpha q$, while coions are modeled as hard spheres with radius $r_i$ and charge $-q$. We also add a symmetric 1:1 salt whose counterions and coions are modeled as hard spheres with radius $r_i$ and charges $\pm q$. In summary, there are three ionic species, bearing charges $+\alpha q$ and $\pm q$ and effective radius $r_I$ and $r_i$, respectively. The $\alpha:1$ salt concentration is $\rho_{\alpha}$ while 1:1 salt concentration is $\rho_1$. The number of ions is defined as $N_{\alpha}=\rho_{\alpha} V$, $N_+=Z+\rho_1 V$ and $N_-=\left[\alpha\rho_{\alpha}+\rho_1\right] V$, where $V=\frac{4\pi}{3}\left[ R^3-a^3 \right]$. We consider the primitive model, in which the dielectric constant of the medium is uniform, $\epsilon$. The Bjerrum length, defined as $\lambda_B=\beta q^2/\epsilon$, is set to $7.2~$\AA, value corresponding to water at room temperature, where $\beta=1/k_BT$, $k_B$ being the Boltzmann constant and $T$ the temperature. The radii of coions, counterions and ions from 1:1 salt is set to $r_i=2~$\AA, while the radius of $\alpha$-valent ions is $r_I=8~$\AA. The SC radius is set to $R=150~$\AA.
%%%%%%%%%%%%%%%% figure %%%%%%%%%%%%%%%%%%%%%
%%%%%%%%%%%%% end of figure %%%%%%%%%%%%%%%%%

\section{The modified Poisson-Boltzmann approach}

\begin{figure}[h]
\begin{center}
\includegraphics[width=5cm]{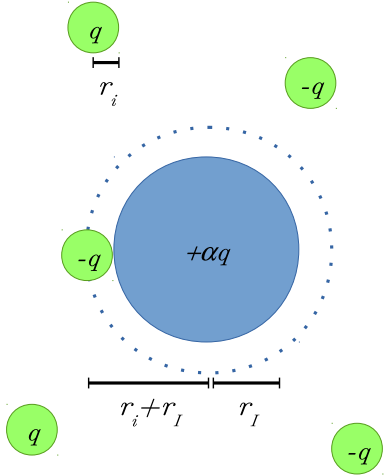}
\end{center}
\caption{A centered ion of charge $+\alpha q$ and radius $r_I$ in an electrolyte solution of ions with charge $\pm q$ and radii $r_i$.}
\label{fig2}
\end{figure}

We intend to use a mean-field theory to study this system. The PB theory can be applied with a very good precision to electrolytes with small ionic radii. If the ions are big enough the PB theory breaks down, as a consequence of the excluded region delimited by ionic particles, which is not taken into account at the traditional PB level. If we want to treat these particles at the PB point-particle level, we must consider that the ionic charge is modified by this exclusion region of radius $r_i+r_I$, see Fig.~\ref{fig2}.
%%%%%%%%%%%%%%%% figure %%%%%%%%%%%%%%%%%%%%%
%%%%%%%%%%%%% end of figure %%%%%%%%%%%%%%%%%

To see how this charge rescaling can be performed, consider a point particle with charge $+\alpha q$ with a spherical exclusion region around it of radius $r_i+r_I$, located at the origin of an electrolyte solution with inverse Debye length $\kappa$, given by $\kappa=\sqrt{4\pi \lambda_B (2\rho_1+\alpha \rho_{\alpha})}$. One can solve the linear PB equation for inside (empty) and for outside (electrolyte) regions considering the standart boundary conditions imposed by Maxwell equations~\cite{Le02}. This results in an electrostatic potential $\Phi(r)$ for the region outside the ionic core which can be written as
%%%%%%%%%%%%%%
\begin{equation}
\Phi(r)=\frac{\alpha q e^{\kappa (r_i+r_I)}}{[1+\kappa (r_i+r_I)]} \frac{e^{-\kappa r}}{\epsilon r} \ ,
\label{lin}
\end{equation}
%%%%%%%%%%%%%%
where $r$ is the distance from the origin. This linearized potential is the basis of the electrostatic part of DLVO potential~\cite{VeOv48}, widely applied for colloidal stability. For a point particle with no exclusion region the solution is the well known Yukawa potential,
%%%%%%%%%%%%%%
\begin{equation}
\Phi(r)=\frac{\alpha q e^{-\kappa r}}{\epsilon r} \ .
\end{equation}
%%%%%%%%%%%%%%
We can then understand that the spherical exclusion region ``changes" the ionic charge by the factor $\dfrac{e^{\kappa (r_i+r_I)}}{[1+\kappa (r_i+r_I)]}$. With this interpretation in mind, our method is based on a mPB equation in which the $\alpha$-valent ionic charge is rescaled by this factor. The prefactor that renormalizes the ionic charge is therefore
%%%%%%%%%%%%%%
\begin{equation}
\theta=\frac{e^{\kappa(r_i+r_I)}}{[1+\kappa(r_i+r_I)]} \ .
\end{equation}
%%%%%%%%%%%%%%
The non-linear mPB equation can be written as
%%%%%%%%%%%%%%
\begin{equation}\label{epb}
\nabla^2 \phi(r)=-\frac{4\pi q}{\epsilon}\left[-Z\delta(r-a)+\alpha\rho_{\alpha}(r)+\rho_+(r)-\rho_-(r)\right] \ ,
\end{equation}
%%%%%%%%%%%%%%
where $\phi(r)$ represents the electrostatic potential around the centered colloid, and $\rho_{\alpha}(r)$ is the $\alpha$-valent local ionic concentration. The ionic distributions read as
%%%%%%%%%%%%%%
\begin{equation}\label{conc1}
\rho_{\alpha}(r)=A_{\alpha}e^{-\beta q \alpha  \theta \phi(r)-\beta U_e(r)} \ ,
\end{equation}
\begin{equation}\label{conc2}
\rho_{\pm}(r)=A_{\pm}e^{\mp \beta q \phi(r)} \ .
\end{equation}
%%%%%%%%%%%%%%
The concept of effective ionic charge is obtained with a linearized PB equation, see Eq.(\ref{lin}), and used in the nonlinear PB equation as a first approximation to define a finite-size correction for the ionic charge.
Notice that the Boltzmann factor corresponding to the $\alpha$-valent ionic distribution has been renormalized via the replacement $\alpha\rightarrow \alpha\theta$. The potential $U_e(r)$, in Eq.(\ref{conc1}), is an exclusion potential which avoids $\alpha$-valent ions to overlap the colloidal core and the SC boundary. The normalization constants are given by
%%%%%%%%%%%%%%
\begin{equation}\label{cons1}
A_{\alpha}=N_{\alpha}/\left[4\pi\int_{a+r_i}^{R-r_i}dr\ r^2 e^{-\beta \alpha q \theta \phi(r)-\beta U_e(r)}\right] \ ,
\end{equation}
\begin{equation}\label{cons2}
A_{\pm}=N_{\pm}/\left[4\pi\int_{a+r_i}^{R-r_i}dr\ r^2 e^{\mp \beta q \phi(r)}\right] \ .
\end{equation}
%%%%%%%%%%%%%%

The Eq.(\ref{epb}) can be solved iteratively together with Eqs.(\ref{cons1}) and (\ref{cons2}), with a mixing parameter to get convergent profiles. It is important to mention that the boundary conditions considered in the linearized PB equation that leads to Eq.(\ref{lin}) have nothing to do with the boundary conditions imposed in the mPB equation, Eq.(\ref{epb}). The linearized PB equation was considered only to construct the concept of  ionic effective charge that incorporate ionic finite size effects.

\section{Monte Carlo Simulations}

%%%%%%%%%%%%%%%% figure %%%%%%%%%%%%%%%%%%%%% revtex figure
% \begin{figure}[t]
% \begin{center}
% \includegraphics[width=7cm]{fig4a.eps}\vspace{0.1cm}
% \includegraphics[width=7cm]{fig4b.eps}\vspace{0.1cm}
% \end{center}
% \caption{Parameters are the same as Fig.~\ref{fig3}, except $r_I=8\AA$.}
% \label{fig4}
% \end{figure}
%%%%%%%%%%%%% end of figure %%%%%%%%%%%%%%%%%

%%%%%%%%%%%%%%%% figure %%%%%%%%%%%%%%%%%%%%% revtex figure
% \begin{figure}[t]
% \begin{center}
% \includegraphics[width=7cm]{fig6a.eps}\vspace{0.1cm}
% \includegraphics[width=7cm]{fig6b.eps}\vspace{0.1cm}
% \end{center}
% \caption{Parameters are the same as Fig.~\ref{fig5}, except $r_I=8\AA$.}
% \label{fig6}
% \end{figure}
%%%%%%%%%%%%% end of figure %%%%%%%%%%%%%%%%%

In order to test our theory we compare our results with MC simulations. The system under study is ilustrated in Fig.~\ref{fig1}. The total energy used in simulations can be written as
%%%%%%%%%%%%%%
\begin{equation}
E=-\sum_{j=1}^N \frac{q_j  Zq}{\epsilon r_j} + \sum_{j=1}^N \sum_{k>j} \frac{q_j  q_k}{\epsilon r_{jk}}  \ ,
\end{equation}
%%%%%%%%%%%%%%
where $N=N_{\alpha}+N_++N_-$, $r_j$ is the distance of ion $j$ from center of the cell, $q_j$ is the charge of ion $j$, $r_{jk}$ is the distance between ions $j$ and $k$. The first term on the right-hand side corresponds to the ionic electrostatic interactions with the centered colloid, whereas the second term is the ion-ion electrostatic interactions. Two types of ionic moves are considered - short and long random moves from previous position. The movements that lead to overlaps between ions, between ions and the colloid and between ions and the limit of SC are rejected. The regular Metropolis algorithm is used with $1\times 10^5$ steps per particle to equilibrate and $1\times 10^3$ steps per particle to get a saved sample. The density profiles are obtained with $3\times 10^5$ samples.

\section{Density Functional Theory}

Apart from the above outlined mPB theory, we also apply a DFT  approach in order to test the accuracy of the new model, as well as to access the relevance of the different ionic correlations to this system. We now briefly summarize the underlying approximations. Taking into consideration hard-core and electrostatic ionic interactions, the total free energy can be split into ideal and mean-field interactions, as well as hard-core and electrostatic correlation contributions. Accordingly, the ionic chemical potentials are separated into these different contributions, and a straightforward application of the Euler-Lagrange condition leads to the following ionic contributions:
\begin{equation}
\rho_{i}(r)=A_{i}e^{-\beta q z_i \phi(r)-\beta\mu_i^{\hc}(r)-\beta\mu_i^{\el}(r)},
\label{EL}
\end{equation}
where $\mu_i^{\hc}(r)$ and $\mu_i^{\el}(r)$ represent local chemical potentials resulting from ionic hard-sphere interactions and electrostatic correlations, respectively. Similarly to the PB approach, the normalizing constants $A_i$ should be calculated so as to ensure the condition of fixed number of ions within the cell (canonical formulation). Notice that when both hard-sphere and electrostatic correlations contributions are neglected, the mean-field ionic profiles are naturally recovered. 

The hard-sphere contributions $\mu_i^{\hc}(r)$ are calculated in the framework of the Rosenfeld's FMT, known to provide a very accurate description of confined hard spheres up to reasonably high packing fractions \cite{Ros89,Roth02,Roth10}. In general lines, the hard-sphere functional $\cF^{\hc}$ is obtained from a local free-energy density $\Psi({\bf r})$ as:
\begin{equation}
\cF^{\hc}[\rho_i({\bf r})]=\int\Psi({\bf r})d{\bf r},
\label{F_hc}
\end{equation}
where $\Psi({\bf r})$ is considered to be a local function
of weighted-densities $n_{\alpha}({\bf r})=\sum_i\int{d{\bf r}'\rho_i({\bf r}')w^{(\alpha)}_i({\bf r}-{\bf r}')}$. The weighted functions $w^{\alpha}_i({\bf r}-{\bf r}')$ represent fundamental measures of the underlying spherical geometry, and read as:
\begin{eqnarray}
w^{(3)}_i({\bf r}) & = & \Theta(a_i-r)\\
w^{(2)}_i({\bf r}) & = & \delta(a_i-r)\\
\mathbold{w}^{(2)}_i({\bf r}) & = & \dfrac{{\bf r}}{r}\delta(a_i-r)= -\nabla w^{(3)}_i({\bf r}),
\label{weights}
\end{eqnarray}
along with the combinations $w^{(1)}_i({\bf r})=w^{(2)}_i({\bf r})/4\pi a_i$, $w^{(0)}_i({\bf r})=w^{(2)}_i({\bf r})/4\pi a_i^2$ and $\mathbold{w}^{(1)}_i({\bf r})=\mathbold{w}^{(2)}_i({\bf r})/4\pi a_i$, where $a_i$ is the radius of the $i-th$ ionic component. The hard-sphere chemical potentials $\mu_{i}^{\hc}({\bf r})$ follows directly from the functional derivative of Eq.(\ref{F_hc}):
\begin{equation} 
\mu^{\hc}_i({\bf r})=\sum_{\alpha}\int{\dfrac{\partial\Psi}{\partial n_{\alpha}({\bf r}')}w_i^{(\alpha)}({\bf r}'-{\bf r})d{\bf r}'},
\label{mu_hc}
\end{equation}
and possess in the present situation radial symmetry. The local free-energy density $\Psi(n_{\alpha})$ can be obtained from different approximations, based on different limiting behaviors for the bulk limit \cite{Roth10}. Here, we adopt the White-Bear functional, in which $\Psi(n_{\alpha})$ is chosen such as to recover the Mansoori-Carnahan-Starling-Leland~(MCSL) equation of state in the limit of bulk concentrations~\cite{Roth02,Yu02}. 

As for electrostatic correlations, $\mu^{\el}_i({\bf r})$, we apply a second-order functional expansion of the residual (over mean-field) electrostatic functional about a reference bulk fluid \cite{Ros93}. There is some freedom in choosing the appropriate bulk fluid around which the perturbation expansion is performed \cite{Yan15}. Here we take the bulk electrolyte to be the neutral electrolyte with concentrations equal to the mean ionic concentrations inside the SC. In this approximation, the chemical potential which accounts for electrostatic correlations reads as:
\begin{equation}
\mu^{\el}_i({\bf r})=-\sum_{j}\int{c_{ij}^{\mathrm{res}}({\bf r}-{\bf r}')\delta\rho_j({\bf r}')d{\bf r}'},
\label{mu_el}
\end{equation}
where $c_{ij}^{\mathrm{res}}({\bf r}-{\bf r}')=c_{ij}({\bf r}-{\bf r}')+z_i z_j\lambda_B/|{\bf r}-{\bf r}'|$ is the residual pair direct correlation function for ions $i$ and $j$ ($c_{ij}({\bf r}-{\bf r}')$ being the corresponding electrostatic direct correlations for the bulk solutions), and $\delta\rho_j({\bf r}')$ is the difference between the inhomogeneous profiles and the bulk ones. In this work, these direct correlations will be computed in the framework of the Mean Spherical Approximation~(MSA), for which closed analytical expressions are available \cite{Blum81}. It is important to notice at this point that $c_{ij}({\bf r}-{\bf r}')$ depends on the size of ions $i$ and $j$, and therefore finite size effects might also have a non trivial influence on the electrostatic correlations, even in situations where the electrostatic couplings are weak. For dilute systems, these contributions from size effects on electrostatic correlations can be even more relevant than the direct hard sphere correlations represented by Eq.(\ref{mu_hc}).

The bulk expansion approximation invoked in Eq.(\ref{mu_el}) will be clearly less accurate as the ionic profiles start to considerably deviate from their bulk regimes (e. g. when they undergo strong variations close to highly charged interfaces). Obviously, the electrostatic and hard-sphere correlations as calculated from Eqs.(\ref{mu_hc}) and (\ref{mu_el}) are treated at different accuracy levels. Quite recently, a DFT approach has been designed which allows one to introduce MSA electrostatic correlations at a level of approximation similar to the FMT for hard-spheres~\cite{RoGi16}. Since our main goal here is only to establish the relative influence of ionic size effects on both hard-sphere and electrostatic correlations, we do not need to go beyond the simple MSA-bulk expansion approximation employed in Eq.(\ref{mu_el}).

\section{Results}

%%%%%%%%%%%%%%%% figure 3 %%%%%%%%%%%%%%%%%%%%% revtex figure
\begin{figure}[h!]
\begin{center}
\includegraphics[scale=0.6]{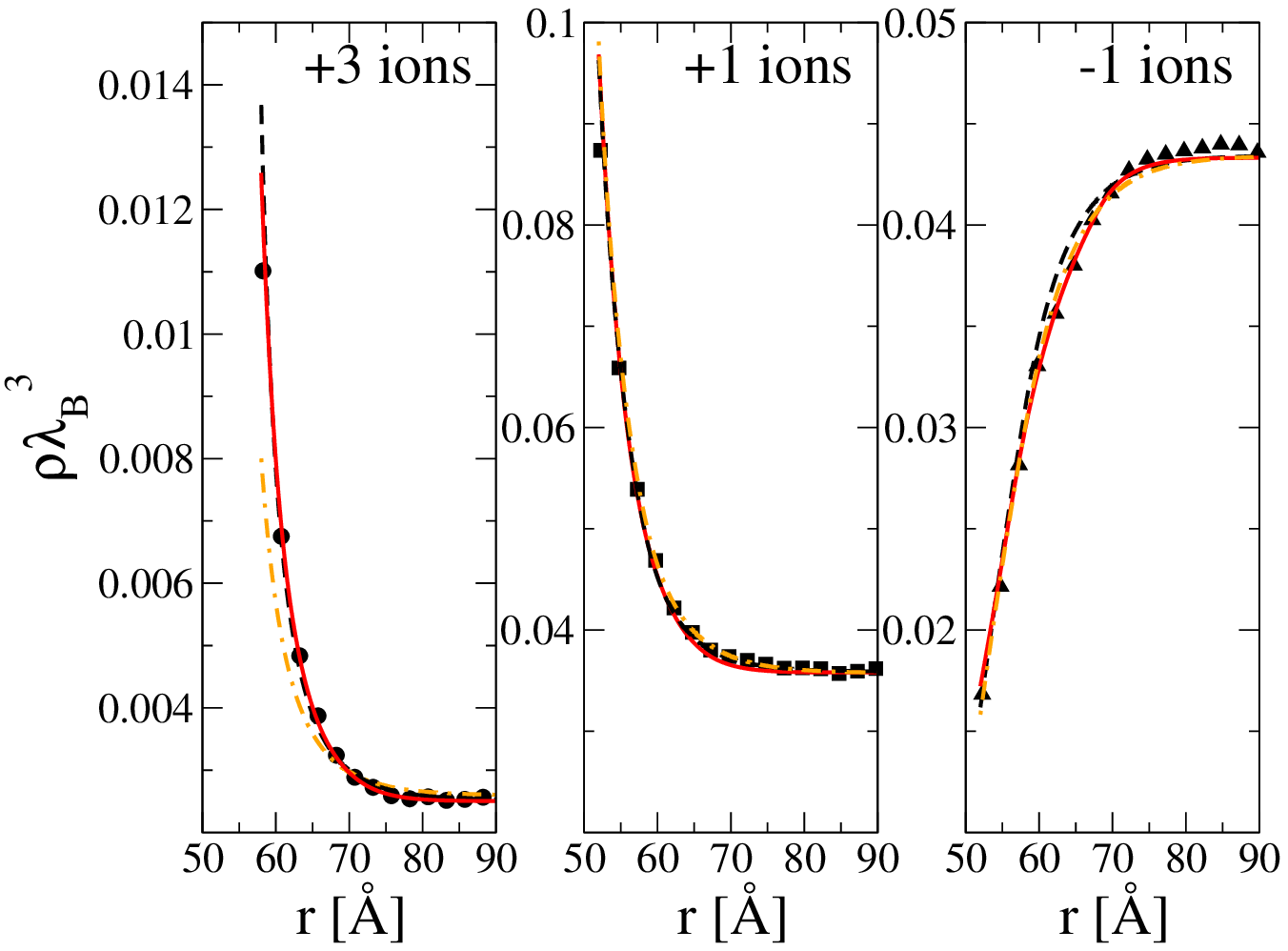}\vspace{0.25cm}

\includegraphics[scale=0.6]{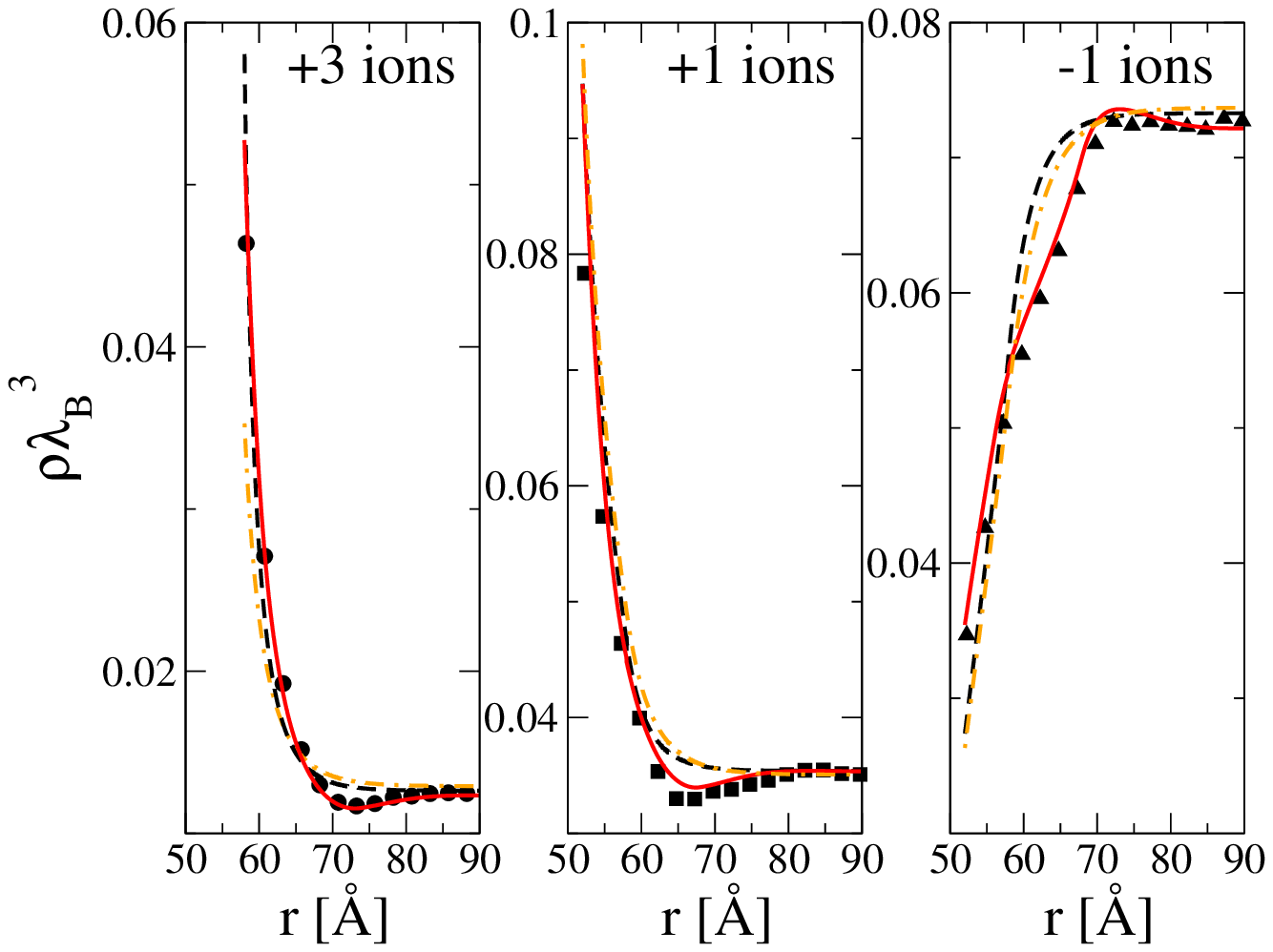}\vspace{0.25cm}
\end{center}
\caption{Ionic density profiles for a mixture of monovalent and trivalent salts. Symbols represent MC simulations data, dashed lines represent the solution of mPB theory while the solid lines are results from the FMT-MSA approach. The dot-dashed lines represent the solution of the traditional PB, setting $\theta=1$. The parameters are $\rho_{\alpha}=10~$mM and $\rho_{\alpha}=50~$mM, for top and bottom figures, respectively.}
\label{fig3}
\end{figure}
%%%%%%%%%%%%% end of figure %%%%%%%%%%%%%%%%%
%%%%%%%%%%%%%%%% figure 4 %%%%%%%%%%%%%%%%%%%%% revtex figure
\begin{figure}[h]
\begin{center}
\includegraphics[scale=0.6]{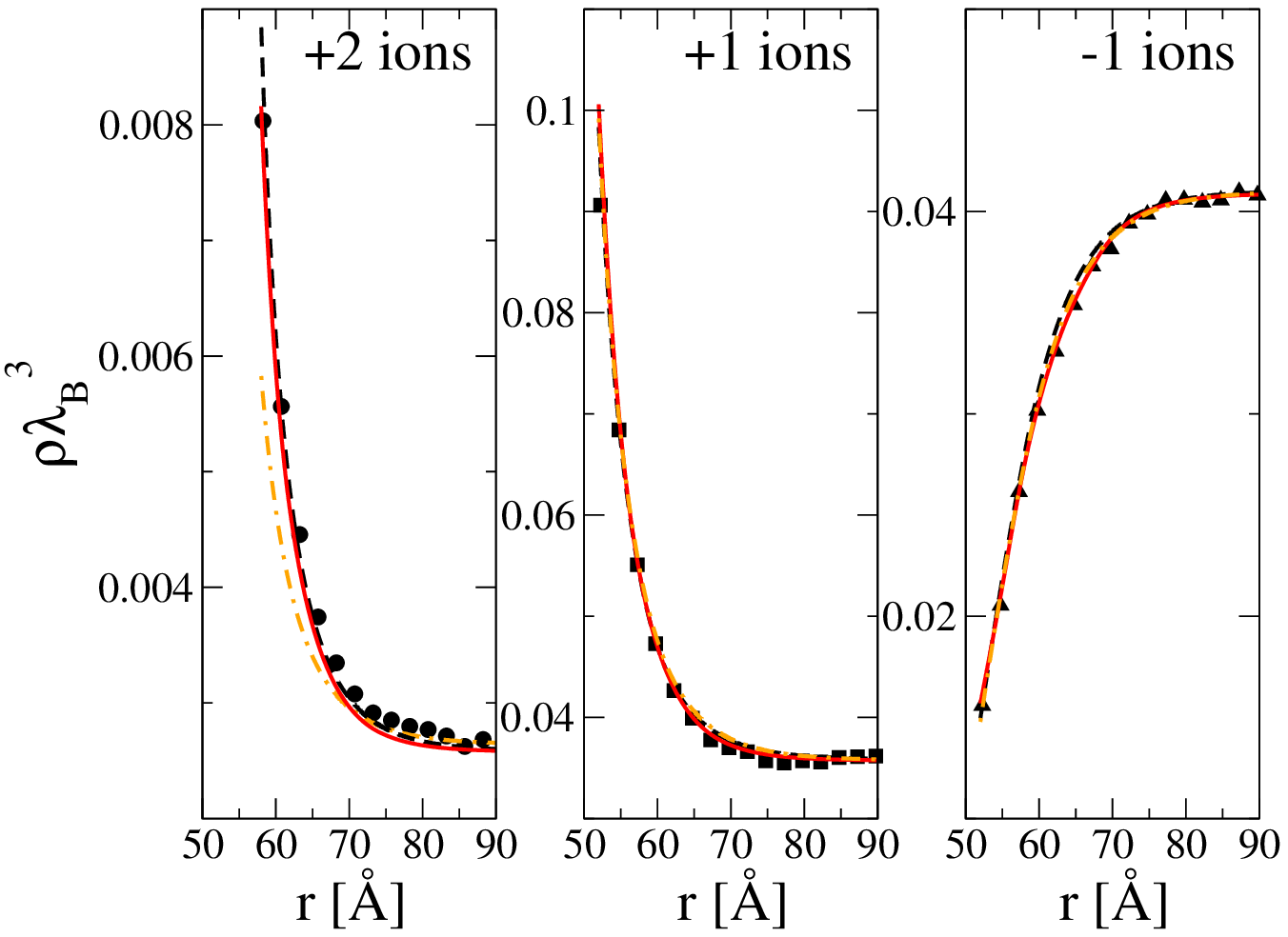}\vspace{0.25cm}

\includegraphics[scale=0.6]{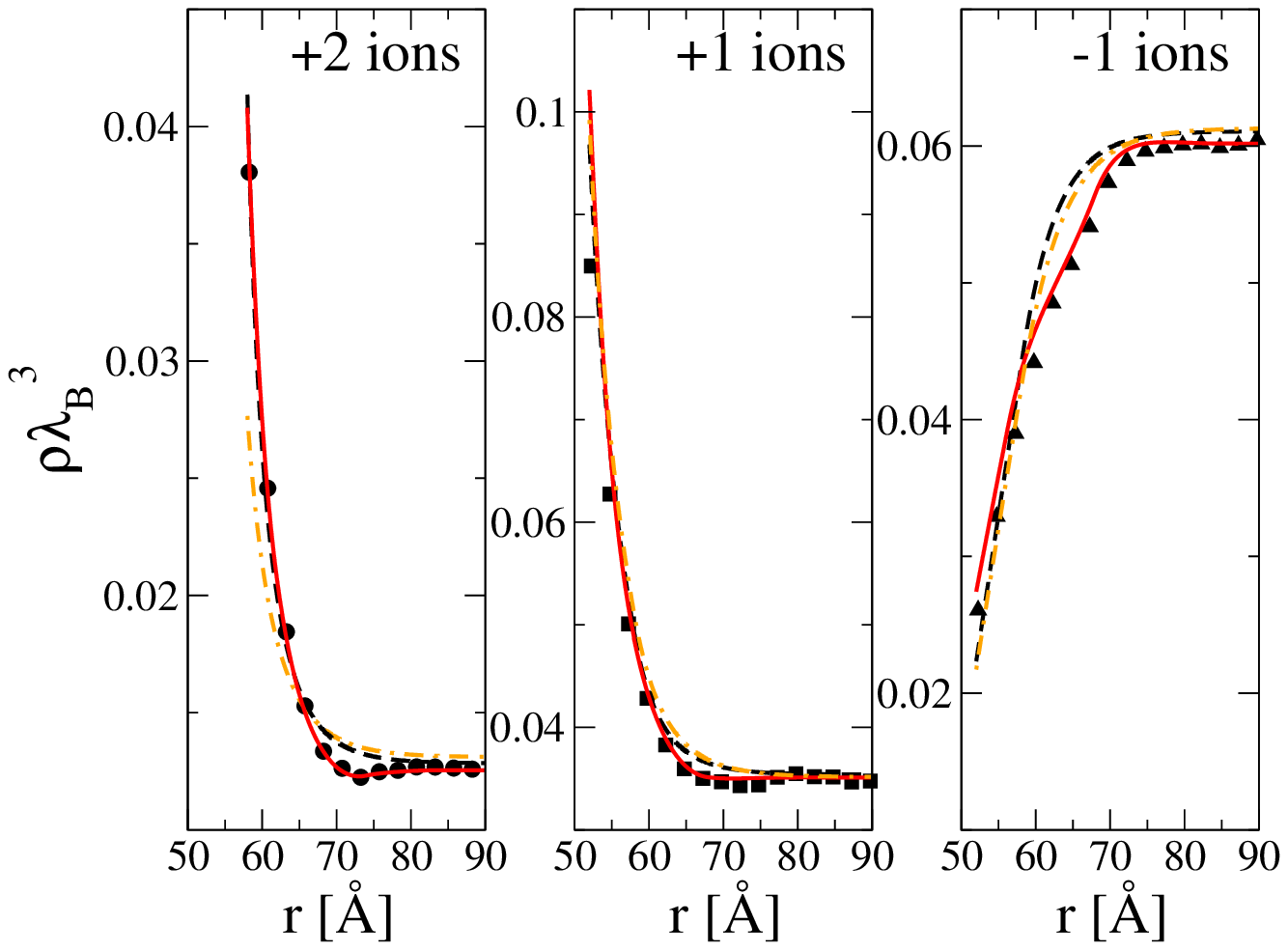}\vspace{0.25cm}
\end{center}
\caption{Ionic density profiles for a mixture of monovalent and divalent salts. Symbols represent MC simulations data, dashed lines represent the solution of mPB theory while the solid lines represent the FMT-MSA theory. The dot-dashed lines represent the solution of the traditional PB, setting $\theta=1$. The parameters are $\rho_{\alpha}=10~$mM and $\rho_{\alpha}=50~$mM, for top and bottom figures, respectively.}
\label{fig4}
\end{figure}
%%%%%%%%%%%%% end of figure %%%%%%%%%%%%%%%%%
We are now going to compare the predictions of the different approaches for describing ionic profiles of the three-component system described in Section~\ref{secII}. The parameters chosen correspond to a colloidal packing fraction of $\eta\approx 0.04$. The concentration of added 1:1 electrolyte is set to be $\rho_1=150$~mM. In Fig.~\ref{fig3} we show density distributions for the case of asymmetric 3:1 electrolyte of concentrations $\rho_{\alpha}=10$~mM (upper panels) and $\rho_{\alpha}=50$~mM (lower panels). Recall that the $\alpha$-valent ions are much larger in size ($r_I=8$~\AA) in comparison with the monovalent ones (for which $r_i=2$~\AA). As we can see, both mPB (dashed curves) and DFT (solid lines) theories describe very well the MC profiles. The agreement is excellent in the case of dilute electrolyte $\rho_{\alpha}=10$~mM, where the density profiles show strict monotonic behaviors. As the concentration becomes larger (bottom curves), the ionic distributions start to present non-monotonic structures close to the colloidal surface, which can not be properly captured by PB-based approaches. Such structures are clearly driven by positional ionic correlations: the high concentration of multivalent counterions electrostatically attached to the charged colloid makes it very unlike to find a neighboring layer of counterions close by, therefore favoring the emergence of a second layer of neutralizing coions. This layering-like structure is a well known feature of electrostatic correlations \cite{Wu11}, and is expected to be more pronounced as the ionic concentrations increase. It is important however to note that the proposed mPB is still able to describe the adsorption of counterions at the colloidal surface very well. 

There are basically three main mechanisms dictating the adsorption of the large $\alpha$-valent counterions onto the colloidal surface. First, their strong electrostatic interactions with the central colloid leads to a large accumulation of these particles close to the colloidal core. This ionic condensation is however limited by the large ionic size, which strongly restrict the number of counterions that can be assembled at the colloidal interface. This effect is further enhanced by the finite surface curvature. Apart from such finite size effects, positional correlations driven by the mutual electrostatic repulsion of neighboring counterions will also limit the degree of counterion association. Neither of these effects are taken into account by the traditional PB approach, which models the ions as point-like particles and completely disregard they electrostatic correlations. However, with the simple ionic charge renormalization proposed in the present mPB approach, size effects can be easily incorporated into the mean-field model, resulting in an accurate description of the adsorption of large ions at the colloidal surface. It is important to notice that, in spite of their higher charge, the electrostatic coupling between trivalent ions $\Gamma_{II}=\alpha^2\lambda_B/2r_I\approx 4$ is not too strong due to their large size. Indeed, this value is compatible with electrolytes made of monovalent ions of size $r_i=0.9$~\AA, for which the PB approach is known to work reasonably well. We therefore expect finite size effects to play the dominant role here, at least for not too high ionic concentrations.  In this limit, it is the colloid-ion correlations -- instead of the ion-ion correlations -- that play the major role. These colloid-ion correlations can be accurately well captured by the proposed  ionic charge rescaling. As the salt concentration becomes larger, ionic positional correlations driven by the presence of $\alpha$-valent ions become the leading contributions, giving rise to the non-monotonic structures observed in Fig.~\ref{fig3}.

Fig.~\ref{fig4} shows the same results for the case of divalent ions, $\alpha=2$. Again, a quite good agreement between both theories and simulations is observed. By decreasing the electrostatic coupling (i. e. the ionic charge) positional correlations driven by electrostatic interactions become weaker. As a consequence, the profiles become less structured in comparison with the trivalent case, as we can see by comparing the lower panels of Figs.~\ref{fig3} and \ref{fig4}. This renders the proposed mPB more accurate for a larger range of ionic concentrations.%\vspace{0.2cm}

\begin{figure}[h]
\begin{center}
\includegraphics[scale=0.6]{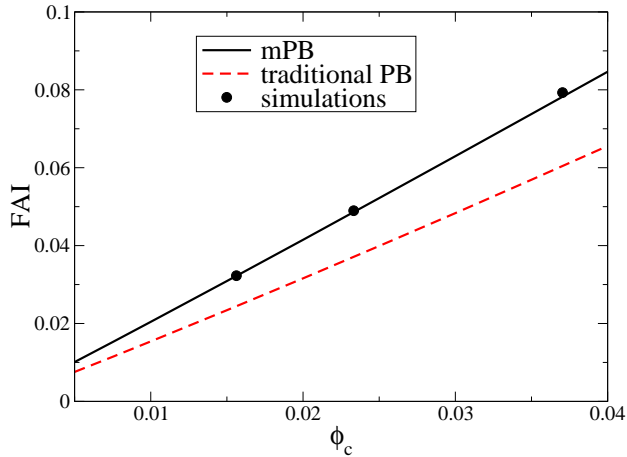}\vspace{0.2cm}
\end{center}
\caption{Fraction of adsorbed ions (FAI) as a function of the colloidal volume fraction, $\phi$. The parameters are the same as in Fig.~\ref{fig3}, top panel, except by the spherical cell radius, $R$.}
\label{fig7}
\end{figure}

Although the proposed mPB predicts ion density profiles that significantly deviates from PB results, the values at the cell boundary -- and therefore the system equation of state -- do not deviate much from their PB counterparts for moderate salt concentrations.  The present method can be also used to obtain the fraction of adsorbed ions~(FAI). We obtain the FAI as a function of the colloidal volume fraction, $\phi_c=a^3/R^3$, for the trivalent salt, see Fig.~\ref{fig7}. This is defined as the fraction of ions which are located at distances less than $2 r_I$ from the colloidal surface. It is obtained from the density profile. The agreement between simulations and mPB solutions is very good, whereas the results from standart PB approach start to considerably deviate from simulation predictions as the confining effects become stronger.
%%%%%%%%%%%%%%%% figure 4 %%%%%%%%%%%%%%%%%%%%% revtex figure
%%%%%%%%%%%%% end of figure %%%%%%%%%%%%%%%%%

\begin{figure}[h]
\begin{center}
\includegraphics[scale=0.6]{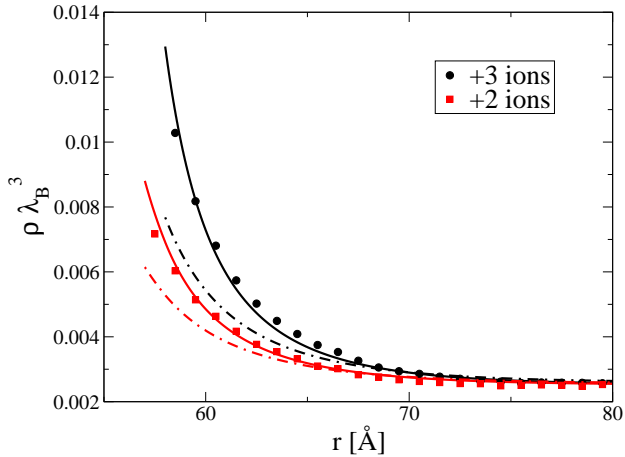}\vspace{0.2cm}
\end{center}
\caption{Density profiles of +3 and +2 ions for a 4 component mixture of monovalent, divalent and trivalent salts. The 2:1 and 3:1 salts are at concentration $0.01~$M. Symbols represent MC simulations data, solid lines represent the solution of mPB equation while dot-dashed lines represent the solution of traditional PB equation.}
\label{fig8}
\end{figure}

It is quite straightforward to extend our model to study electrolytes with more ionic components. We mix 1:1, 2:1 and 3:1 salts considering a four component system. The parameters are the same as in Fig.~\ref{fig3} (top panel), however with the addition of 2:1 salt at concentration $0.01~$M. The effective +2 radius is $7~$\AA, different from +3 ionic radius, which is $8~$\AA. The $\kappa$ parameter must be updated in order to consider the stronger electrolyte, $\kappa=\sqrt{4\pi \lambda_B (2\rho_1+3 \rho_{3}+2 \rho_{2})}$. Apart from the trivalent ions, the bivalent big ions also have their charges renormalized using the proposed Ansatz. Also, the mPB equation, Eq.(\ref{epb}), must be corrected in order to include one more ionic component. The agreement between mPB and simulations is still very good, as can be seen in Fig.~\ref{fig8}.
%%%%%%%%%%%%%%%% figure 4 %%%%%%%%%%%%%%%%%%%%% revtex figure
%%%%%%%%%%%%% end of figure %%%%%%%%%%%%%%%%%

%%%%%%%%%%%%%%%% figure 5 %%%%%%%%%%%%%%%%%%%%% revtex figure
\begin{figure}[h]
\begin{center}
\includegraphics[scale=0.6]{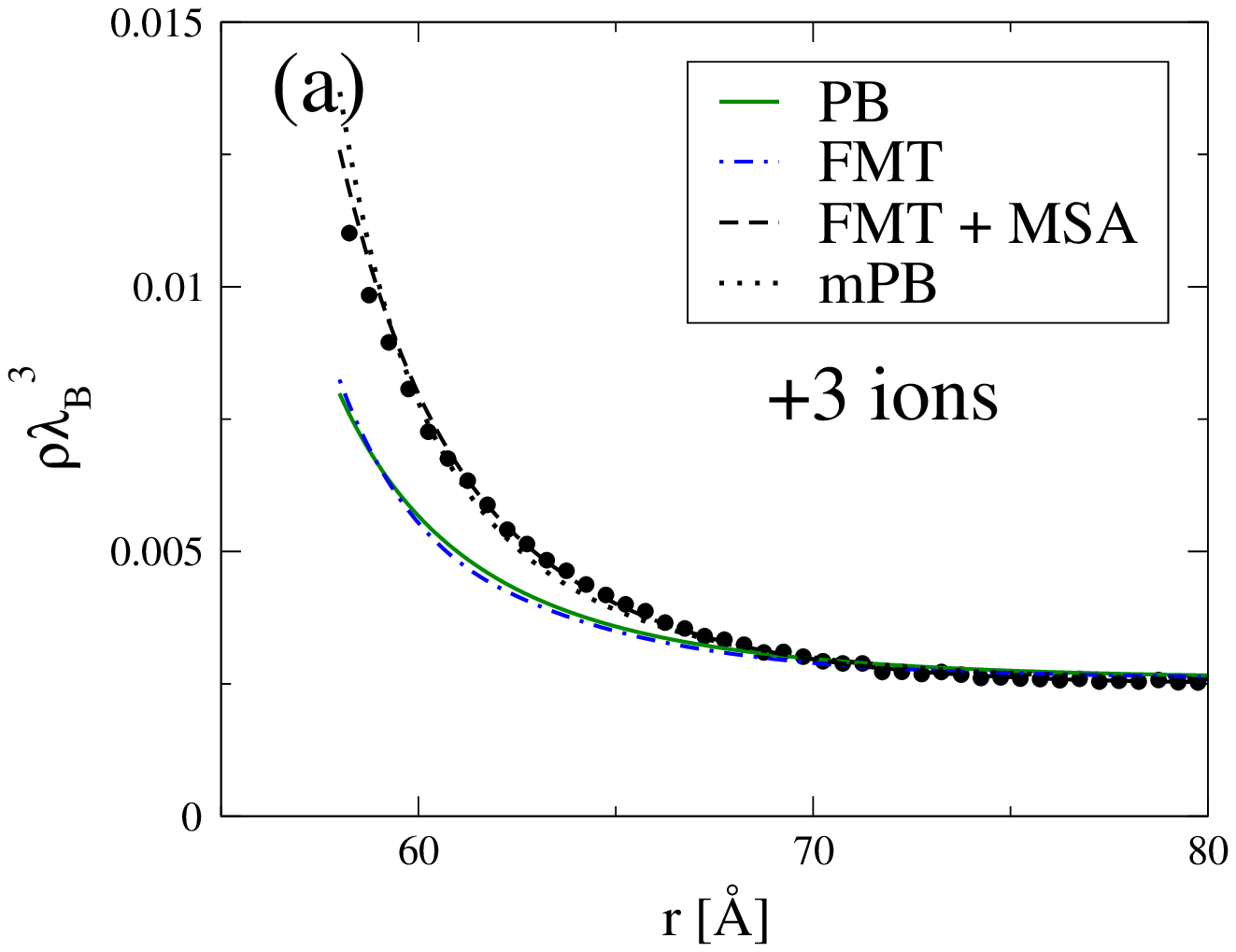}\vspace{0.2cm}
\includegraphics[scale=0.6]{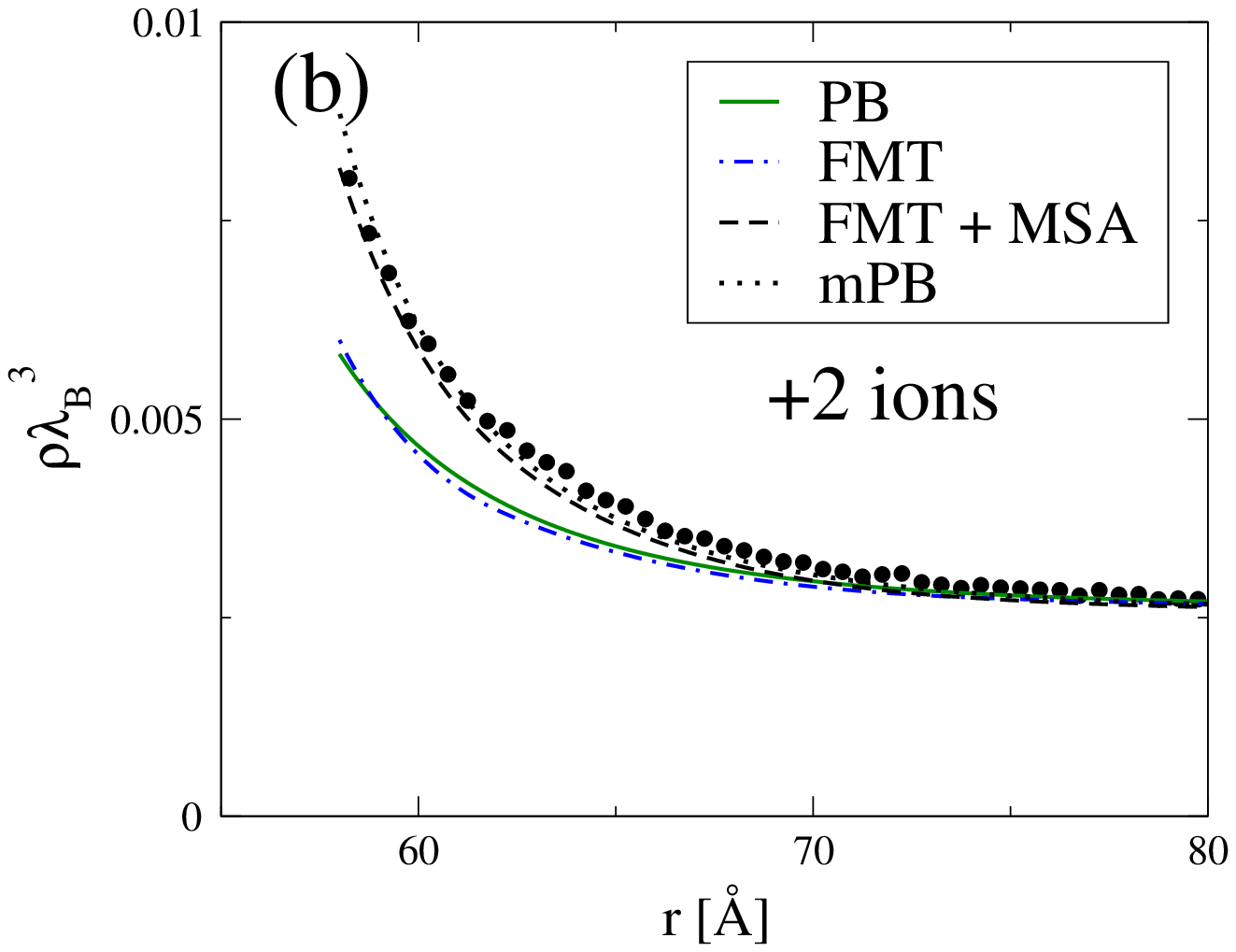}\vspace{0.2cm}
\end{center}
\caption{Ionic distributions for trivalent (a) and divalent (b) ions at the smallest concentration of $\alpha:1$ electrolyte $\rho_{\alpha}=10$~mM. Symbols represent MC simulations data, the solid curves are results from the traditional PB approach, while dot-dashed and dashed curves contain additional effects from size (FMT) and electrostatic (FMT+MSA) effects, respectively. The dot curves represent the results of mPB approach.}
\label{fig5}
\end{figure}
%%%%%%%%%%%%% end of figure %%%%%%%%%%%%%%%%%

In order to perform a deeper analysis on the interplay between electrostatic and size correlations in our thee-component system, we show in Fig.~\ref{fig5} the profiles of $\alpha$-valent ions for the dilute concentration $\rho_{\alpha}=10$~mM. In the framework of the DFT formalism, the effects from different correlations on the density profiles can be conveniently ``triggered" by setting either $\mu_i^{\hc}(r)$ or $\mu_i^{\el}(r)$ to be zero in Eq.(\ref{EL}). This allows one to remove the electrostatic correlations and check the effects of hard-sphere interactions separately. In the absence of any correlations ($\mu_i^{\hc}(r)=\mu_i^{\el}(r)=0$), the PB profiles are recovered. In Fig.~\ref{fig5}, it can be clearly observed that the traditional PB theory strongly underestimates the adsorption of multivalent ions. Interesting enough, the scenario does not change much when only hard-sphere correlations are taken into account through the FMT approach. It is only when the electrostatic correlations are ``switched on" that the counterions are pushed toward the colloidal surface, thereby reproducing the higher adsorption predicted by the simulation results. We can therefore conclude that it is the size effects on the electrostatic correlations -- rather than the direct hard-sphere correlations -- that play the major role in the dilute regime. 

The different mechanisms that lead to finite size effects can be interpreted as follows. Lets assume that a cavity is created around a point-like particle, see Fig.~\ref{fig2}. As the cavity is created, the neighboring particles have to be ``expelled out'' from the cavity region. The work required for this particle re-arrangement is closely related to the hard-core chemical potential $\mu_i^{\hc}(r)$ given by Eq.(\ref{mu_hc}), and leads to hard-core positional correlations. On the other hand, in an electrolyte solution the emergence of a cavity also removes {\it charge carries} away from the cavity volume around the point charge -- which requires additional work to be performed against electrostatic forces. Since these charges are responsible for electrostatic screening, the electrostatic potential around the centered ion, in Fig.~\ref{fig2}, becomes less screened at a given position beyond the cavity region, leading to an enhancement of electrostatic interactions. In order to replace a finite-sized charge by a point charge this charge must be enhanced such as to take into account the loss in screening caused by the void. At the MSA level of approximation, for instance, the point-like charges can be interpreted as spherical shells in which the charge is effectively smeared out over the shell surface \cite{Wei87,RoGi16}. At low ionic concentrations, such screening mechanism dominates over the  hard-core  effects. In particular, it renders the electrostatic interactions between the colloid and  $\alpha$-valent ions stronger, which explains the higher ionic adsorption of these ions as their finite size are properly taken into account. Since the PB approach is unable to distinguish electrostatic interactions between point-like or finite size objects, it can not describe such enhancement of counterions adsorption resulting from size effects. However, we show that a simple rescaling of the ionic charges is able to recover the PB accuracy, whose numerical implementation is way much simpler than the bulk expansions applied for the electrostatic correlations.

\section{Conclusions}

In this work we developed a simple theory based on a correction to PB equation to properly quantify the concentration of ions with high effective radius near a spherical colloid/nanoparticle. The ionic charge considered in the PB theory takes into account the excluded region around the bigger ion. This is accomplished by identifying an effective charge comparing the solution of linear PB equation with and without an exclusion region around the ionic charge. We show that the theory can be applied also for multivalent ions, as their radii are big enough to avoid electrostatic ion-ion correlations. Besides, we applied a DFT to investigate both size and electrostatic effects, and investigate the interplay between these effects. When the ionic concentration becomes sufficiently large, non-monotonic behaviors typical of electrostatic correlation effects start to emerge, leading to the breakdown of mean-field based approaches. In this regime, the proposed DFT based on FMT and bulk-MSA expansion is still able to accurately predict the MC ionic profiles.

The ionic profiles obtained with the proposed mPB approach compares very well with MC simulations for the different combinations of ionic size and charge asymmetries considered, remarkably improving over PB results. The adopted concept of renormalizing charges in order to partially incorporate size and/or nonlinear effects in a simple and intuitive way has been widely applied to a number of  charged systems. The theory breaks down in the regime from moderate to high ionic concentrations, when packing effects resulting from strong size and electrostatic correlations lead to layering structures at the vicinity of the charged interface. The net adsorption of counterions onto the colloidal surface can be still captured to a reasonable degree of accuracy by the present mPB approach. This is to be contrasted to traditional mPB approaches based on either local functional approximations or lattice-based models, which typically fail to predict the ionic concentrations at the contact with a charged surface. While the lattice-based approaches predict ionic profiles that saturates close to the charged interface at large ionic concentrations~\cite{Fry12,GiCo17}, local approximations tend to overestimate the ionic adsorption, as the (unbound) local ionic chemical potentials become increasingly large close to highly charged surfaces~\cite{Ant05,Le02}.

Finally, we point out that the applied SC model can be used as a basis to investigate a number of important properties of concentrated system of nanoparticles, such as the nanoparticle renormalized charges and osmotic properties. The models outlined in this work can therefore be applied to a number of potential applications involving nanoparticles coexisting with polydisperse electrolytes.

\section{Acknowledgments}
This work was partially supported by the CNPq, CAPES and Alexander von Humboldt Foundation.

%\bibliography{ref.bib}

%
\end{document}